\documentclass[hyper]{JHEP} 

\usepackage{epsfig}

\newcommand\fverb{\setbox\pippobox=\hbox\bgroup\verb}

\newcommand\fverbdo{\egroup\medskip\noindent%

            \fbox{\unhbox\pippobox}\ }

\newcommand\fverbit{\egroup\item[\fbox{\unhbox\pippobox}]}

\newbox\pippobox

\title{Note About Hamiltonian Formalism
of Modified $F(R)$ Ho\v{r}ava-Lifshitz Gravities and
Their Healthy Extension}
\author{J. Kluso\v{n}\\
Department of
Theoretical Physics and Astrophysics\\
Faculty of Science, Masaryk University\\
Kotl\'{a}\v{r}sk\'{a} 2, 611 37, Brno\\
Czech Republic\\
E-mail: \email{klu@physics.muni.cz}}
\preprint{\hepth{1002.4859}}

 \abstract{This note is devoted to the
study of  Hamiltonian formalism of
modified F(R) Ho\v{r}ava-Lifshitz theories of gravity
that were proposed recently in 
arXiv:1001.4102[hep-th].
We also study  Hamiltonian formulation
of the healthy extended Ho\v{r}ava-Lifshitz
gravities  and show that
these theories have many unusual and interesting
 properties.}

\def\mH{\mathcal{H}}

\def\partt{\partial_t}

\def\bx{\mathbf{x}}
\def\by{\mathbf{y}}

\newcommand{\mG}{\mathcal{G}}

\newcommand{\bT}{\mathbf{T}}

\def\pb #1{\left\{#1\right\}}

\begin{document}
\section{Introduction and Summary}\label{first}

Last year  Petr Ho\v{r}ava proposed new
intriguing approach for the formulation
of UV finite  quantum theory of gravity
\cite{Horava:2009uw,Horava:2008ih,Horava:2008jf}
.
The basic idea of this theory is to
modify the UV behavior of the general
theory so that the theory is
perturbatively renormalizable. However
this modification is only possible on
condition when we abandon
 Lorentz symmetry in the high energy regime: in
this context, the Lorentz symmetry is
regarded as an approximate symmetry
observed only at low energy.

In
\cite{Kluson:2009rk,Kluson:2009xx} we
introduced version of
 Ho\v{r}ava-Lifshitz gravity that is
related to $F(R)$ theories
\footnote{For review and extensive list
of references, see
\cite{Capozziello:2009nq,Sotiriou:2008rp,Nojiri:2008nt,Faraoni:2008mf,Nojiri:2006ri}.}.
This approach was further developed in
very interesting paper
\cite{Chaichian:2010yi}. It was argued there
that such a form of gravity could provide
 unification
of the early time inflation with the
late time acceleration. Moreover, the
preliminary analysis of the cosmological
solution  with some promising
properties was given there as well.

The goal of this short note is to find
the  Hamiltonian formulation of modified
$F(R)$ Ho\v{r}ava-Lifshitz theory. In fact,
 the Hamiltonian analysis of given theory
 was already done in \cite{Chaichian:2010yi}
but we feel that it deserve to be investigated
further. Following
\cite{Deruelle:2009pu} we formulate
the Hamiltonian formalism for modified
$F(R)$ Ho\v{r}ava-Lifshitz gravity and
we show that the algebra of constraints
is closed for theory that obeys the
projectability condition that claims that
the lapse function depends on time only
 $N=N(t)$.

As a counterexample of  standard form of
well defined 
Hamiltonian dynamics of  modified $F(R)$ 
theories of gravity that obey the projectability
 condition we  discuss the Hamiltonian
 analysis of  healthy extended
  Ho\v{r}ava-Lifshitz gravity that
 was proposed in 
\cite{Blas:2009qj,Blas:2009ck}.
 Explicitly, since the momentum conjugate to
 lapse is primary constraint of the theory
 we find that the  preservation of this constraint
 during the time evolution of the system
 induces the secondary constraint that
 has non-zero Poisson bracket with the primary
 constraint $p_N
\approx 0$. In other words, they form the
collection of the second class  constraints.
It is instructive to compare this result with 
conclusions presented in
\cite{Henneaux:2009zb}. It was shown there
that the Ho\v{r}ava-Lifshitz gravity without
the projectability condition has very peculiar
property in the sense that the Hamiltonian
constraints are the second class constraints and
that the gravitational Hamiltonian vanishes strongly.
However in case of the healthy extended Ho\v{r}ava-Lifshitz
gravities we find new and surprasing resolution.
Explicitly, since $p_N$ and corresponding 
secondary constraints are the second class constraints
their can be explicitly solved. Then we can
express $N$ as a function of cannonical variables,
at least at principle. Further,  the reduced phase
space of healthy extended Ho\v{r}ava-Lifshitz
theory is spanned by $g_{ij},p^{ij}$ and there
is no gauge freedom related to the time reparameterization
of theory since there is not the first class Hamiltonian
constraint. Interestingly, this result naturally solves
the problem of the closure of the algebra of  the Hamiltonian
constraints in the Ho\v{r}ava-Lifshitz gravity. Secondly,
one can hope that heatlhy extended Ho\v{r}ava-Lifshitz
gravities can provide  solution of  the problem of time in gravity
\footnote{For  detailed discussion of this problem, see 
\cite{Isham:1992ms}.}. We hope to return to this interesting
problem in future. 

The structure of this note is as follows. In
the next section (\ref{second}) we perform the
Hamiltonian analysis of modified $F(R)$ Ho\v{r}ava-Lifshitz
theory of gravity. In section (\ref{third}) we
perform the Hamiltonian analysis of healthy extended
Ho\v{r}ava-Lifshitz gravities and discuss their
properties.



\section{Hamiltonian Formulation of
Modified $F(R)$ Ho\v{r}ava-Lifshitz
gravity}\label{second}

Let us consider $D+1$ dimensional
manifold $\mathcal{M}$ with the
coordinates $x^\mu \ , \mu=0,\dots,D$
and where $x^\mu=(t,\bx) \ ,
\bx=(x^1,\dots,x^D)$. We presume that
this space-time is endowed with the
metric $\hat{g}_{\mu\nu}(x^\rho)$ with
signature $(-,+,\dots,+)$. Suppose that
$ \mathcal{M}$ can be foliated by a
family of space-like surfaces
$\Sigma_t$ defined by $t=x^0$. Let
$g_{ij}, i,j=1,\dots,D$ denotes the
metric on $\Sigma_t$ with inverse
$g^{ij}$ so that $g_{ij}g^{jk}=
\delta_i^k$. We further introduce the operator
$\nabla_i$ that is covariant derivative
defined with the metric $g_{ij}$. 
 We  introduce  the
future-pointing unit normal vector
$n^\mu$ to the surface $\Sigma_t$. In
ADM variables we have
$n^0=\sqrt{-\hat{g}^{00}},
n^i=-\hat{g}^{0i}/\sqrt{-\hat{g}^{
00}}$. We also define  the lapse
function $N=1/\sqrt{-\hat{g}^{00}}$ and
the shift function
$N^i=-\hat{g}^{0i}/\hat{g}^{00}$. In
terms of these variables we write the
components of the metric
$\hat{g}_{\mu\nu}$ as
\begin{eqnarray}
\hat{g}_{00}=-N^2+N_i g^{ij}N_j \ ,
\quad \hat{g}_{0i}=N_i \ , \quad
\hat{g}_{ij}=g_{ij} \ ,
\nonumber \\
\hat{g}^{00}=-\frac{1}{N^2} \ , \quad
\hat{g}^{0i}=\frac{N^i}{N^2} \ , \quad
\hat{g}^{ij}=g^{ij}-\frac{N^i N^j}{N^2}
\ .
\nonumber \\
\end{eqnarray}
Then it is easy to see that
\begin{equation}
\sqrt{-\det \hat{g}}=N\sqrt{\det g} \ .
\end{equation}
We further define the extrinsic
derivative
\begin{equation}
K_{ij}=\frac{1}{2N}
(\partial_t g_{ij}-\nabla_i N_j-
\nabla_j N_i) \ .
\end{equation}
It is well known that the components of
the Riemann tensor can be written in
terms of ADM variables \footnote{For
review and extensive list of
references, see
\cite{Gourgoulhon:2007ue}.}. For
example, in case of Riemann curvature
we have
\begin{equation}\label{R}
R=K^{ij}K_{ij}-K^2+R^{(D)}+\frac{2}{\sqrt{-\hat{g}}}
\partial_\mu(\sqrt{-\hat{g}}n^\mu K)-
\frac{2}{\sqrt{g}N}\partial_i
(\sqrt{g}g^{ij}\partial_j N) \ ,
\end{equation}
where $K=K_{ij}g^{ji}$ and where
$R^{(D)}$ is Riemann curvature
calculated using the metric $g_{ij}$.
  The new
formulation of Ho\v{r}ava-Lifshitz
 $F(R)$ gravity that was given
in \cite{Chaichian:2010yi}
 is based on the modification of the
relation (\ref{R}). In fact, the action
introduced there
 takes the form
\begin{equation}\label{actionNOJI}
S_{F(\tilde{R})}= \int dt d^D\bx
\sqrt{g}N F (\tilde{R}) \ ,
\end{equation}
where
\begin{equation}
\tilde{R}= K_{ij}\mG^{ijkl}K_{kl}+
\frac{2\mu}{\sqrt{-\hat{g}}}
\partial_\mu (\sqrt{-\hat{g}}n^\mu K)
-\frac{2\mu}{\sqrt{g}N}
\partial_i (\sqrt{g}g^{ij}\partial_j N)
-E^{ij}\mG_{ijkl}E^{kl} \ ,
\end{equation}
where $\mu$ is constant  and where the
generalized metric $\mG^{ijkl}$  is
defined as
\begin{equation}
\mG^{ijkl}=\frac{1}{2}(g^{ik}g^{jl}+
g^{il}g^{jk})-\lambda g^{ij}g^{kl} \ ,
\end{equation}
where $\lambda$ is real constant.
$E^{ij}$ are defined using the
variation of 
$D-$dimensional  action $W(g_{kl})$
\begin{equation}
\sqrt{g}E^{ij}=\frac{\delta W}{\delta
g_{ij}} \ .
\end{equation}
These objects were introduced in the
original work \cite{Horava:2009uw}.
However we can consider theory when
$E_{ij}\mG^{ijkl}E_{kl}$ is replaced
with more general terms  that depend on
$g_{ij}$ and their covariant
derivatives. Further, the action
(\ref{actionNOJI})
 is invariant under foliation
preserving diffeomorphism
\begin{equation}
t'-t=f(t) \ , \quad
x'^i-x^i=\xi^i(t,\bx) \ .
\end{equation}

 Our goal
is to perform the detailed Hamiltonian
analysis of the theory defined by the
action (\ref{actionNOJI}).
 In
order to do this  we
 introduce two non-dynamical
fields $A,B$ and
 rewrite the
  action (\ref{actionNOJI})
into the form
\begin{equation}
S_{F(\tilde{R})}= \int dt d^D\bx
\sqrt{g}N (B(\tilde{R}-A)+F(A)) \ .
\end{equation}
It is easy to see that solving the
equation of motion with respect to
$A,B$ this action reduces into
(\ref{actionNOJI}). On the other hand
when we  perform  integration by parts
we obtain the action in the form
\begin{eqnarray}\label{SFtR}
S_{F(\tilde{R})}
=\int dt d^D\bx \left( \sqrt{g}N B(
K_{ij}\mG^{ijkl}K_{kl}
-E^{ij}\mG_{ijkl}E^{kl}-A)+\nonumber
\right.
\\
\left. +\sqrt{g}N F(A) -2\mu
\sqrt{g}(\partial_t B-N^i\partial_i B)
K  + 2\mu
\partial_i B \sqrt{g}g^{ij}
\partial_j N \right) \ , \nonumber \\
\end{eqnarray}
where we ignored the boundary terms.
From this form of the action we clearly
see that $B$ is now dynamical field. In
fact, from the action (\ref{SFtR}) we
find the conjugate momenta
%
\begin{eqnarray}\label{defmom}
p_N&=&\frac{\delta
S_{F(\tilde{R})}}{\delta
\partt N}\approx 0 \ , \quad
p_i=\frac{\delta
S_{F(\tilde{R})}}{\delta
\partt N^i} \approx 0 \ , \quad
p_A=\frac{\delta
S_{F(\tilde{R})}}{\delta
\partt A}\approx 0 \ , \nonumber \\
p^{ij}&=&\frac{\delta
S_{F(\tilde{R})}}{\delta
\partt
g_{ij}}=\sqrt{g}\left(B\mG^{ijkl}K_{kl}-
\frac{2\mu  g^{ij}}{N}(\partt B-N^i
\partial_i B)\right) \ , 
\nonumber \\
\pi&=&\frac{\delta
S_{F(\tilde{R})}}{\delta \partt B}=
-2\mu\sqrt{g}K \ . \nonumber \\
\end{eqnarray}
The first line in (\ref{defmom})
implies that $p_N,p_i$ are primary
constraints of the theory. On the other
hand the relations on the second and
third line in (\ref{defmom}) can be
inverted so that
\begin{eqnarray}
& &(\partt B-N^i\partial_i B) =
-\frac{N}{2\mu D\sqrt{g}}
\left(\frac{1}{2\mu}B(1-\lambda
D)\pi+p^{ij}g_{ji}\right) \ ,  \nonumber \\
& &K_{ij}=\frac{1}{B\sqrt{g}}
\mG_{ijkl}\left( p^{kl}-\frac{1}{D}
g^{kl}\left(\frac{1}{2\mu}B(1-\lambda
D)\pi+ p^{kl}g_{lk}\right)\right) \ ,
\nonumber \\
\end{eqnarray}
where we used the fact that
\begin{equation}
g_{ij}\mG^{ijkl}=(1-\lambda D)g^{kl} \
.
\end{equation}
Using these results it is
straightforward exercise to find
corresponding Hamiltonian
\begin{equation}
H=\int d^D\bx (N \mH_T+N^i\mH_i+v^A
p_A+v^N p_N+v^ip_i ) \ ,
\end{equation}
where
\begin{eqnarray}
\mH_T&=& \frac{1}{B\sqrt{g}}p^{ij}
 \mG_{ijkl}p^{kl}
 -\frac{1}{D\mu\sqrt{g}} (1-\lambda D)^2 \pi
 p^{ij}g_{ji}-
\frac{1}{B D\sqrt{g}}(1-\lambda D)
(\pi^{ij}g_{ji})^2+
 \nonumber \\
&+&\frac{1}{\sqrt{g}}\frac{(1-\lambda
D)^2 B}{4 D\mu^2}
((1-\lambda D)^2-2)\pi^2 +\nonumber \\
 &+&\sqrt{g}B(E^{ij}\mG_{ijkl}E^{kl}
+A)-\sqrt{g} F(A)+ 2\mu \partial_i[
\partial_j B \sqrt{g}
 g^{ij}] \ , \nonumber \\
\mH_i&=& -2  g_{ik}\nabla_j p^{kj} +
\pi\partial_i B \ , \nonumber \\
\end{eqnarray}
and where we included the primary
constraints $p_N\approx 0 \ ,
p_i\approx 0 \ , p_A\approx 0$.
Note that as opposite to the
Hamiltonian analysis presented in
\cite{Chaichian:2010yi} we find that
$B$ is dynamical field. 
Further, the  consistency of the primary
 constraints
with the time evolution of the system
implies following secondary constraints
\begin{eqnarray}
\partial_t p_N(\bx)&=&
\pb{p_N(\bx),H}=-\mH_T(\bx)\approx 0 \
,
\nonumber \\
\partial_t p_i(\bx)&=&\pb{
p_i(\bx),H}=-\mH_i(\bx)\approx 0 \ ,
\nonumber \\
\partial_t p_A(\bx)&=&
\pb{p_A(\bx),H}=-\sqrt{g}N(B-F'(A))(\bx)\equiv
-\sqrt{g}NG_A(\bx)\approx 0 \ . \nonumber
\\
\end{eqnarray}
Since $\pb{p_A(\bx),G_A(\by)}=
F''(A)\delta(\bx-\by)$ we see that
$(p_A,G_A)$ are the second class
constraints and hence can be explicitly
solved. The solving the first one we
set $p_A$ strongly zero while
  solving the second one
 we find
$F'(A)=B$. If we presume that
$F'$ is invertible we can express
$A$ as  a function of $B$ so that
 $A=\Psi(B)$ for some function $\Psi$.
Finally, since
$\pb{p^{ij},p_A}=\pb{g_{ij},p_A}=0$ we
see that the Dirac brackets between
canonical variables coincide with
Poisson brackets.

Let us  consider the smeared form of
the spatial diffeomorphism generator
\begin{equation}
\bT_S=\int d^D\bx \xi^i \mH_i \ .
\end{equation}
It is easy to see that this generates
the spatial diffeomorphism  since
\begin{eqnarray}
\pb{\bT_S,B(\bx)}&=&-\xi^i(\bx)\partial_i B(\bx) \ , \nonumber \\
\pb{\bT_S,\pi(\bx)}&=&-\xi^i(\bx) \partial_i \pi(\bx)-
\partial_i \xi^i (\bx)\pi(\bx) \ , \nonumber \\
\pb{\bT_S,g_{ij}(\bx)}&=&
-\xi^k(\bx)\partial_k
g_{ij}(\bx)-g_{jk}(\bx)
\partial_k \xi^k(\bx)
-g_{ik}(\bx)\partial_j\xi^k(\bx) \ ,
\nonumber \\
\pb{\bT_S,p^{ij}(\bx)}&=& -\partial_k
p^{ij}(\bx) \xi^k(\bx)-p^{ij}(\bx)\partial_k
\xi^k(\bx)+p^{jk}(\bx)
\partial_k \xi^i(\bx)+p^{ik}(\bx)
\partial_k \xi^j(\bx) \ . \nonumber \\
\end{eqnarray}
Using the Poisson bracket between
$\bT_S$ and $B$ we find
 \begin{eqnarray}
 \pb{\bT_S,A(B(\bx))}&=&\frac{\delta A(\bx)}{\delta B(\bx)}
 \pb{\bT_S,B(\bx)}=\nonumber \\
&=&
 -\frac{\delta A(\bx)}{\delta B(\bx)}\xi^k(\bx)
 \partial_k B(\bx)=
 -\xi^k(\bx) \partial_k A(\bx) \ . \nonumber \\
 \end{eqnarray}
Then we find following Poisson
bracket  
\begin{equation}
\pb{\bT_S,\mH_T(\bx)}=
-\xi^k(\bx)\partial_k \mH_T(\bx)-
\mH_T(\bx)\partial_k \xi^k(\bx)
\end{equation}
that implies
\begin{equation}\label{STpb}
\pb{\bT_S(\xi),\bT_T(f)}= \int d^D\bx
(\partial_k f \xi^k)\mH_T=
\bT_T(\partial_k f \xi^k ) \ . 
\end{equation}
Note that the right side in the
expression above vanishes for constant $f$.

Finally we calculate the Poisson
bracket of $\bT_T(f),\bT_T(g)$. Clearly
the calculations of the Poisson bracket
$\pb{\mH_T(\bx),\mH_T(\by)}$ will be 
as intricate as the calculation of
the Poisson bracket  between these
constraints in standard Ho\v{r}ava-Lifshitz
gravity. The structure of these
brackets was analyzed in 
\cite{Li:2009bg,Henneaux:2009zb} with 
the outline that $\mH_T$ are the second
class constraints with unclear physical
meaning of this theory. On the other
hand it is possible to find
consistent physical theory (at least
on the classical level) in case when we 
 impose the projectability condition
that claims that $N=N(t)$. Then the
local primary constraint
$p_N(\bx)\approx 0$ is replaced with
the global one $p_N\approx 0$ and its
preservation during the time evolution
of the system implies the global
constraint \footnote{Clearly this
constraint takes the same form as
$\bT_T(f)$ for constant $f=1$.}
\begin{equation}
\bT=\int d^D\bx \mH_T(\bx)\approx 0 \ .
\end{equation}
Then we find that the Hamiltonian is
the linear combination of the first
class constraints
\begin{equation}
H=v^N p_N+v^i p_i + N \bT+\bT_S(N^i) \
.
\end{equation}
Finally  using the fact that
\begin{equation}
\pb{\bT_S(\xi),\bT_S(\eta)}=
\bT_S(\xi^i\partial_i\eta^k -\eta^i
\partial_i \xi^k)
\end{equation}
and also the equation (\ref{STpb}) when
we impose the condition $f=1$ we find
that the constraints $ \bT\approx 0  \
, \bT_S(\xi)\approx 0$ are consistent
with the time evolution of the system
since
\begin{eqnarray}
\partial_t \bT&=&\pb{\bT,H}\approx 0 \ , \nonumber \\
\partial_t \bT_S(\xi)&=&\pb{\bT_S(\xi),H}
 \approx 0 \ .  \nonumber \\
\end{eqnarray}
Let us conclude our results. We derived
the Hamiltonian formulation of modified
$F(R)$ Ho\v{r}ava-Lifshitz gravity and
argued that this is a consistent theory 
when the projectability condition is
imposed. Observe that the requirement of
the  consistency
of the constraints with the time evolution
implies the secondary constraints only 
which is different from analysis presented
 in \cite{Chaichian:2010yi}. Explicitly, it
was argued there  the consistency of the
constraints with the time evolution of
the system could lead to the the possibility of the generation of
tertiary constraints or constraints of 
higher order until the closure of constraints 
is established.

\section{ Hamiltonian
Dynamics of Healthy Extended 
Ho\v{r}ava-Lifshitz Gravity}\label{third}
The  healthy extended
of  Ho\v{r}ava-Lifshitz theory was proposed
in \cite{Blas:2009qj} in order to improve
some pathological properties of the
Ho\v{r}ava-Lifshitz gravity without projectability
condition. Explicitly, the
healthy extended Ho\v{r}ava-Lifshitz
gravity is the version  the Ho\v{r}ava-Lifshitz
theory without projectability
and without detailed balance condition
imposed 
that contains additional vector
$a_i$ constructed from
the lapse function $N(t,\bx)$ as
\begin{equation}
a_i=\frac{\partial_i N}{N}
\end{equation}
Note that under foliation preserving
diffeomorphism where $N'(t',\bx')=
N(t,\bx)(1-\dot{f}(t))$ we find that
$a_i$ transforms as
\begin{equation}
a'_i(t',\bx')=a_i(t,\bx)-a_j(t,\bx)
\partial_i \xi^j(t,\bx) \ .
\end{equation}
Let us now consider the
healthy extension of
modified $F(R)$ Ho\v{r}ava-Lifshitz
theory of gravity
defined by the action
\begin{eqnarray}
S=\int dt d^D\bx \sqrt{g}N (B(\tilde{R}
-V(g_{ij},a_i)-A)+F(A)) \ , 
\end{eqnarray}
where $V(g,a)$ is an additional
potential term that depends on
$a_i$ and on $g_{ij}$. Performing the same
analysis as in previous section
we find the Hamiltonian in
the form
\begin{eqnarray}
H&=&\int d^D\bx
\left(N(\mH_T+B\sqrt{g}V)+N^i\mH_i
+\right. \nonumber \\
&+&v^i p_i+v^N p_N+v^A p_A) \ , \nonumber \\
\nonumber \\
\end{eqnarray}
where $\mH_T$ and $\mH_i$ are the 
same as in case of modified $F(R)$
Ho\v{r}ava-Lifshitz gravity.
The crucial point of the Hamiltonian
analysis of the healthy extended
Ho\v{r}ava-Lifshitz gravity is
that the condition of the preservation
of the primary constraint $p_N\approx 0$
implies following secondary one
\begin{eqnarray}
\partial_t p_N(\bx)&=&
\pb{p_N(\bx),H}=
-(\mH_T(\bx)+B\sqrt{g}V(\bx))+\nonumber \\
&+&
\frac{1}{N}\partial_i\left(NB\frac{\delta
V}{\delta a_i}\right)(\bx)\equiv
-\tilde{\mH}_T(\bx)\approx 0
\nonumber \\
\end{eqnarray}
using
\begin{eqnarray}
\pb{p_N(\bx),\int d^D\by N B
\sqrt{g}V(g,a)}= -B\sqrt{g}V(\bx)
+\frac{1}{N}
\partial_i \left(NB\sqrt{g}\frac{\delta V}{\delta a_i}
\right)(\bx)
\nonumber \\
\end{eqnarray}
The general analysis of the constraint
systems implies that the total Hamiltonian
is the sum of the original Hamiltonian
and all constraints so that 
the Hamiltonian takes the form
\begin{equation}\label{Hamhealt}
H=\int d^D\bx(
N(\mH_T+\sqrt{g}BV)+N^i\mH_i
+v_T\tilde{\mH}_T +v^N p_N+v^ip_i) \ ,
\end{equation}
where $v_T$ is Lagrange multiplier related
to the new constraint $\tilde{\mH}_T$. Observe
that as opposite to the case of canonical
 gravity or 
 standard Ho\v{r}ava-Lifshitz
theory $N$ does not appear as Lagrange
multiplier in the Hamiltonian (\ref{Hamhealt}).  
This is the first indication  of
the slightly unusual behavior of
this theory. In order to investigate
the properties of given theory further 
we introduce
the
smeared form of the Hamiltonian
constraint $\bT_T(f)=\int d^D\bx
f(\bx)\mH_T(\bx)$. Then we find
\begin{eqnarray}
\pb{p_N,\bT_T(f)}&=&
\frac{1}{N}f\partial_i \left(B\sqrt{g}
\frac{\delta V}{\delta a_i}\right)
+\nonumber \\
&+&\partial_i\left(\frac{f}{N}\right)\frac{\partial_j N}{N}
B\sqrt{g}\frac{\delta^2 V}{\delta
a_i a_j}+
\partial_j\left(\partial_i
\left(\frac{f}{N}\right)B\sqrt{g}\frac{\delta^2
V}{\delta a_i \delta a_j}\right) \ . 
\nonumber \\
\end{eqnarray}
Since the Hamiltonian can be written as
\begin{equation}\label{Hheathy}
H=\int d^D\bx (
N(\mH_T+\sqrt{g}BV)+v^Np_N+v^ip_i)+
\bT_T(v_T)+\bT_S(N^i)
\end{equation}
we find that the time derivative of
$p_N$ is equal to
\begin{eqnarray}
\partt p_N&=&
\pb{p_N,H}\approx 
\frac{1}{N}v_T\partial_i \left(B\sqrt{g}
\frac{\delta V}{\delta a_i}\right)
+\nonumber \\
&+&\partial_i\left(\frac{v_T}{N}\right)\frac{\partial_j N}{N}
B\frac{\delta^2 V}{\delta
a_i a_j}+
\partial_j\left(\partial_i
\left(\frac{v_T}{N}\right)B\sqrt{g}\frac{\delta^2
V}{\delta a_i \delta a_j}\right) \ . 
\nonumber \\
\nonumber\\
\end{eqnarray}
In principle this equation can be solved for $v_T$
so that it  is determined 
by the dynamical variables. 
In other words, $p_N$
and $\tilde{\mH}_T$ form the second
class constraints 
and consequently  there is no gauge freedom
related to the constraint
$\tilde{\mH}_T$. However this fact
has very interesting consequences
for the structure of the theory
\footnote{I would like to thank 
to Diego Blas, Oriol Pujolas and Sergey Sibiryakov
for suggesting me this interpretation.}.
Explicitly, since $p_N(\bx),\tilde{\mH}_T(\bx)$
are second class constraints they can be explicitly
solved. The solution of the first one is $p_N(\bx)=0$
strongly. On the other hand we suggest that
the constraint $\tilde{\mH}_T(\bx)=0$ can be
solved for $a_i=\frac{\partial_i N}{N}$ and hence 
 $N$ can be expressed as  a function
of dynamical variables $g_{ij},p^{ij}$ 
\begin{equation}\label{Ncan}
N=\Phi(g_{ij},p^{ij}) \ . 
\end{equation}
Further, since the Poisson brackets
between $g_{ij},p^{ij}$ and $p_N$ vanish
we find that the Dirac brackets between
cannonical variables $g_{ij},p^{ij}$
that span the reduced phase
space of the theory coincide with the 
Poisson brackets. Finally, using 
(\ref{Ncan} in (\ref{Hheathy}) we 
find that  the Hamiltonian on the reduced
phase space takes  the form
\begin{equation}
H=\int d^D\bx (
\Phi(\mH_T+\sqrt{g}BV(\Phi))+v^ip_i)+
+\bT_S(N^i) \ . 
\end{equation}
We see that this Hamiltonian contains
generator of the spatial diffeomorphism 
that is the first class constraint. The presence
of this constraint is a consequence
of the fact that this theory is 
invariant under spatial diffeomorphism.
Observe that  there is no gauge
freedom related to time reparameterization.
This result suggests that even if
the structure of the  healthy extended
Ho\v{r}ava-Lifshitz gravity is
  completely different from
 general relativity it has the
potential that it can solve
 the long standing problem 
of time in general relativity
\footnote{By "the problem of time" in General Relativity (GR) one means that GR is a completely parametrised
system. That is, there is no natural notion of time due to the diffeomorphism invariance of the theory
and therefore the canonical Hamiltonian which generates time reparametrisations vanishes. In fact,
instead of a Hamiltonian there are an infinite number of spatial diffeomorphism and Hamiltonian
constraints respectively, of which the canonical Hamiltonian is a linear combination, which generate
infinitesimal spacetime diffeomorphisms.}. It would be
very interesting to study this 
theory further for some  
examples of the potential $V$ that allow 
to find   $N$ as a function of 
 cannonical variables and hence
find Hamiltonian on reduced phase space.
 We hope to retun to this
problem in future.

\vskip 5mm

 \noindent {\bf
Acknowledgements:}
I would like to thank 
to Diego Blas, Oriol Pujolas and Sergey Sibiryakov
for comments considering the first version
of my paper and for suggestion of the
correct interpretation of results derived here.
 This work   was
supported by the Czech Ministry of
Education under Contract No. MSM
0021622409. \vskip 5mm



\begin{thebibliography}{20}







\bibitem{Horava:2009uw}
  P.~Horava,
\emph{``Quantum Gravity at a Lifshitz
Point,''}
  Phys.\ Rev.\  D {\bf 79} (2009) 084008
  [arXiv:0901.3775 [hep-th]].



\bibitem{Horava:2008ih}
  P.~Horava,
\emph{``Membranes at Quantum
Criticality,''}
  JHEP {\bf 0903} (2009) 020
  [arXiv:0812.4287 [hep-th]].

\bibitem{Horava:2008jf}
  P.~Horava,
\emph{``Quantum Criticality and
Yang-Mills Gauge Theory,''}
  arXiv:0811.2217 [hep-th].


















\bibitem{Kluson:2009rk}
  J.~Kluson,
\emph{``Horava-Lifshitz f(R)
Gravity,''}
  arXiv:0907.3566 [hep-th].







\bibitem{Blas:2009qj}
  D.~Blas, O.~Pujolas and S.~Sibiryakov,
\emph{``A healthy extension of Horava
gravity,''}
  arXiv:0909.3525 [hep-th].



\bibitem{Blas:2009ck}
  D.~Blas, O.~Pujolas and S.~Sibiryakov,
\emph{``Comment on `Strong coupling in
extended Horava-Lifshitz gravity',''}
  arXiv:0912.0550 [hep-th].






\bibitem{Li:2009bg}
  M.~Li and Y.~Pang,
\emph{``A Trouble with
Ho\v{r}ava-Lifshitz Gravity,''}
  JHEP {\bf 0908} (2009) 015
  [arXiv:0905.2751 [hep-th]].

 
\bibitem{Chaichian:2010yi}
  M.~Chaichian, S.~Nojiri, S.~D.~Odintsov, M.~Oksanen and A.~Tureanu,
\emph{``Modified F(R) Horava-Lifshitz
gravity: a way to accelerating FRW
cosmology,''}
  arXiv:1001.4102 [hep-th].


\bibitem{Henneaux:2009zb}
  M.~Henneaux, A.~Kleinschmidt and G.~L.~Gomez,
\emph{``A dynamical inconsistency of
Horava gravity,''}
  arXiv:0912.0399 [hep-th].





\bibitem{Papazoglou:2009fj}
  A.~Papazoglou and T.~P.~Sotiriou,
\emph{``Strong coupling in extended
Horava-Lifshitz gravity,''}
  arXiv:0911.1299 [hep-th].

\bibitem{Kluson:2009xx}
  J.~Kluson,
\emph{``New Models of f(R) Theories of
Gravity,''}
  arXiv:0910.5852 [hep-th].




\bibitem{Gourgoulhon:2007ue}
  E.~Gourgoulhon,
\emph{``3+1 Formalism and Bases of
Numerical Relativity,''}
  arXiv:gr-qc/0703035.


\bibitem{Capozziello:2009nq}
  S.~Capozziello, M.~De Laurentis and V.~Faraoni,
\emph{``A bird's eye view of
f(R)-gravity,''}
  arXiv:0909.4672 [gr-qc].

\bibitem{Sotiriou:2008rp}
  T.~P.~Sotiriou and V.~Faraoni,
\emph{``f(R) Theories Of Gravity,''}
  arXiv:0805.1726 [gr-qc].

\bibitem{Nojiri:2008nt}
  S.~Nojiri and S.~D.~Odintsov,
\emph{``Dark energy, inflation and dark
matter from modified F(R) gravity,''}
  arXiv:0807.0685 [hep-th].

\bibitem{Faraoni:2008mf}
  V.~Faraoni,
\emph{``f(R) gravity: successes and
challenges,''}
  arXiv:0810.2602 [gr-qc].


\bibitem{Nojiri:2006ri}
  S.~Nojiri and S.~D.~Odintsov,
\emph{``Introduction to modified
gravity and gravitational alternative
for dark energy,''}
  eConf {\bf C0602061} (2006) 06
  [Int.\ J.\ Geom.\ Meth.\ Mod.\ Phys.\  {\bf 4} (2007) 115]
  [arXiv:hep-th/0601213].

\bibitem{Deruelle:2009pu}
  N.~Deruelle, Y.~Sendouda and A.~Youssef,
\emph{``Various Hamiltonian formulations of f(R)
gravity and their canonical
relationships,''}
  Phys.\ Rev.\  D {\bf 80} (2009) 084032
  [arXiv:0906.4983 [gr-qc]].


\bibitem{Isham:1992ms}
  C.~J.~Isham,
\emph{``Canonical quantum 
gravity and the problem of time,''}
  arXiv:gr-qc/9210011.










\end{thebibliography}
\end{document}